\documentclass[aps,prl,english,superscriptaddress,twocolumn,floatfix]{revtex4}
\usepackage[T1]{fontenc}
\usepackage[latin9]{inputenc}
\usepackage[a4paper]{geometry}
\usepackage{bm}
\usepackage{amstext}
\usepackage{graphicx}
\usepackage{color}

\makeatletter
\@ifundefined{textcolor}{}
{%
 \definecolor{BLACK}{gray}{0}
 \definecolor{WHITE}{gray}{1}
 \definecolor{RED}{rgb}{1,0,0}
 \definecolor{GREEN}{rgb}{0,1,0}
 \definecolor{BLUE}{rgb}{0,0,1}
 \definecolor{CYAN}{cmyk}{1,0,0,0}
 \definecolor{MAGENTA}{cmyk}{0,1,0,0}
 \definecolor{YELLOW}{cmyk}{0,0,1,0}
 }

\makeatother

\usepackage{babel}

\begin{document}

\title{Monopoles, magnetricity and the stray field from spin ice}

\author{Stephen~J.~Blundell}
\altaffiliation{Email: s.blundell@physics.ox.ac.uk}
\affiliation{Oxford University Department of Physics, 
Clarendon Laboratory, Parks Road, Oxford OX1 3PU, United Kingdom}

\date{\today}

\begin{abstract}
An analysis is presented of the behaviour of muons in the
low-temperature state in spin ice.  It is shown in detail how the
behavior observed in some previous muon experiments on spin ice in a
weak transverse field may result from the macroscopic stray field of
magnetized spin ice.  A model is presented which allows these
macroscopic field effects to be simulated and the results agree with
experiment.  The persistent spin dynamics at low temperature originate
from the sample and could be a muon-induced implantation effect that
is operative in out-of-equilibrium systems with long relaxation times.
\end{abstract}

\pacs{999}

\maketitle 

Spin ice \cite{harris,gingras,jaubert2011} is the name given to
compounds such as Dy$_2$Ti$_2$O$_7$ in which the magnetic Dy ions
($\mu\approx 10\mu_{\rm B}$) sit on a pyrochlore lattice (composed of
corner-sharing tetrahedra) and are constrained by easy-axis anisotropy
only to point in or out of each tetrahedron.  The long-ranged dipolar
interactions are almost perfectly screened at low temperatures
\cite{hertog,isakov} so that the low-energy properties are essentially
identical to an effective frustrated nearest-neighbour model
equivalent to Pauling's model for proton disorder in water ice
\cite{pauling}.  The effective exchange energy in this dipolar Ising
magnet is minimized if the moments on each tetrahedron satisfy the ice
rules, namely that two spins point in and two spins point out
\cite{harris}. This results in a state whose large degeneracy is
quantitatively consistent with the measured thermodynamic properties
\cite{ramirez}.  Very recently it was realised that the natural
excitations \cite{ryzhkin} of spin ice can be described as magnetic
monopoles \cite{castelnovo2008}: flipping a moment breaks the ice
rules in two neighbouring tetrahedra (one tetrahedron has moments
arranged three-out, one-in; its neighbor has one-out, three-in) and
these states are positive and negative monopoles.  Further flips allow
these monopoles individually to hop through the lattice. This
deconfined monopole picture is an economical description of the
low-temperature behavior of spin ice, as one has moved from
considering a concentrated collection of localised dipolar spins to a
more dilute array of itinerant particles interacting Coulombically, a
so-called magnetic Coulomb liquid \cite{castelnovo2008}.

\begin{figure}
\includegraphics[width=8.1cm]{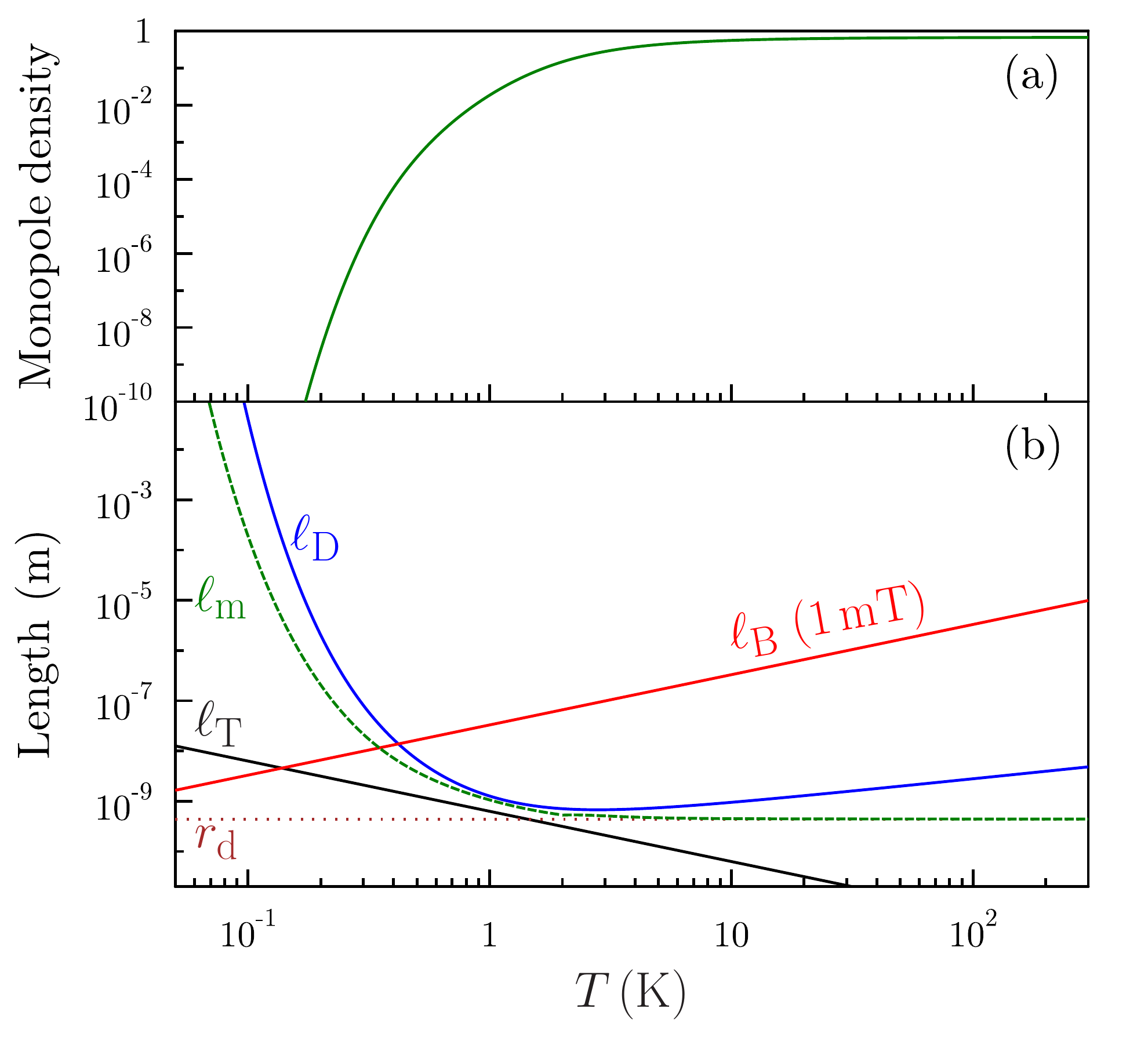}
\caption{\label{lengths-muons} 
(a) The monopole density (in units of monopoles per tetrahedron)
 within the framework of Debye-H\"uckel theory \cite{castelnovo}.
(b) The key lengthscales in spin ice.
}
\end{figure}

The number of monopoles in spin ice can be estimated within the
framework of Debye-H\"uckel (DH) theory \cite{castelnovo}, see
Fig.~\ref{lengths-muons}(a), and at very low temperatures the
monopoles are very sparse indeed.  It has been argued
\cite{bramwell,giblin} that spin ice should conduct magnetic charge
(``magnetricity'') in an analogous manner to an electrolyte and that
Onsager's theory \cite{onsager} of the Wien effect should apply to
spin ice.  In this context, it is useful to consider the key
lengthscales: the lattice spacing $a\approx 10$\,\AA\ (so that the
distance from the centre of one tetrahedron to the centre of its
neighbour is $r_{\rm d}=\sqrt{3}a/4\approx 4.4$\,\AA, and the monopole
charge is $Q=2\mu/r_{\rm d}=4.50\mu_{\rm B}$\AA$^{-1}$), the Bjerrum
length $\ell_{\rm T}=\mu_0Q^2/(8\pi k_{\rm B}T)$ (the distance below
which monopoles are bound by the Coulomb interaction), the field
length $\ell_{\rm B}=k_{\rm B}T/QB$ (the lengthscale above which
drift of monopoles in a field is discernable), the mean minimum distance
between monopoles $\ell_{\rm m}$ (related to $n_{\rm f}$, the number
density of free monopoles) and the Debye length $\ell_{\rm D}=(k_{\rm
  B}T/2\mu_0Q^2n_{\rm f})^{1/2}$ (the distance above which monopole
charge density is screened) \cite{giblin}.  The temperature
dependences of these lengthscales are plotted in
Fig.~\ref{lengths-muons}(b).

\begin{figure}
\includegraphics[width=8.1cm]{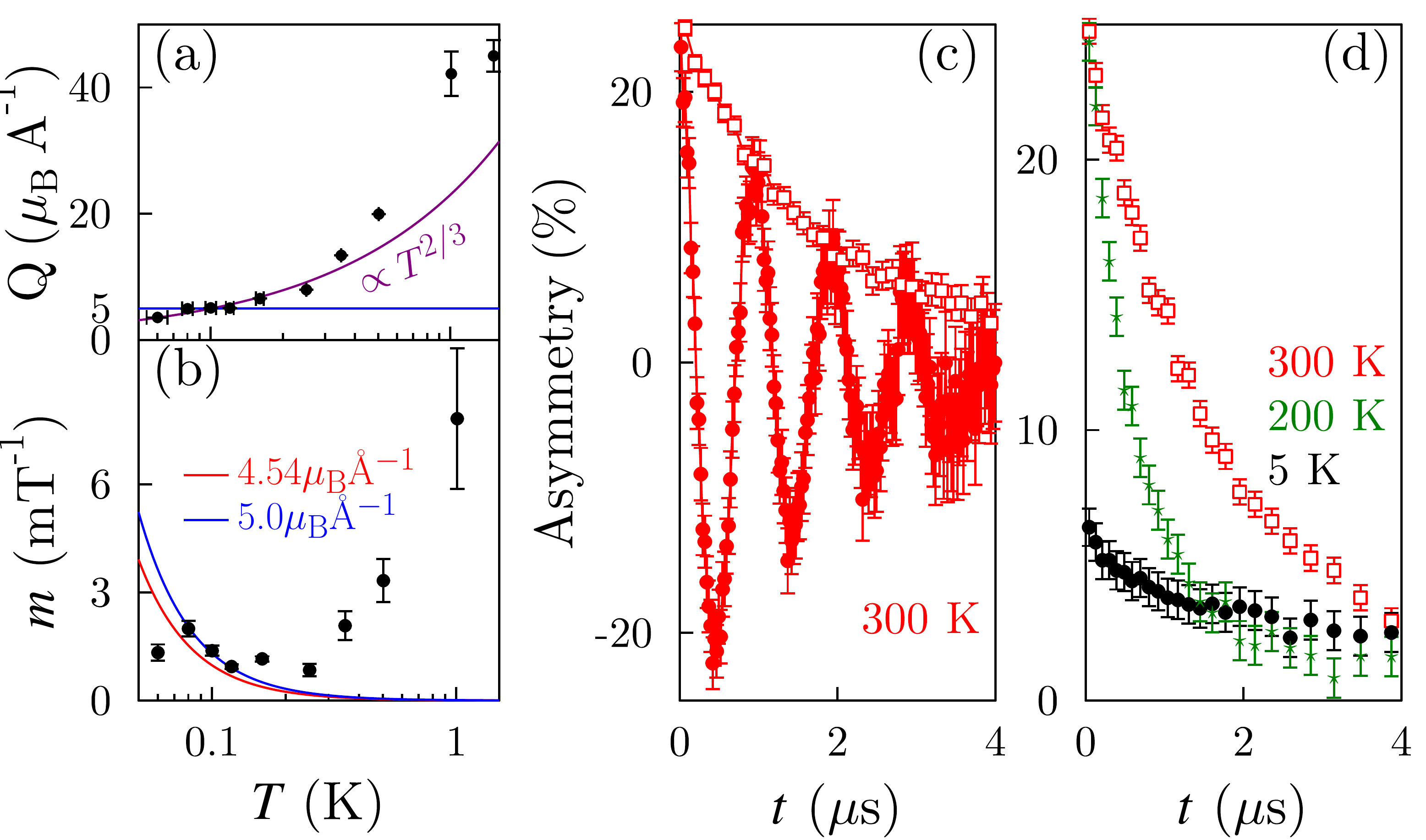}
\caption{\label{muondata} (a) The extracted monopole charge $Q$ from
  the data in Ref.~\onlinecite{bramwell} showing the plateau close to
  $Q=5\mu_{\rm B}$\AA$^{-1}$, close to the theoretical value
  4.54\,$\mu_{\rm B}$\AA$^{-1}$ \cite{castelnovo2008}.  The
  low-$T$ data roughly follow a $T^{2/3}$ dependence.  (b) The
  same data transformed to values of $m={1 \over \lambda(0)} {{\rm
      d}\lambda(B) \over {\rm d}B}$ in order to test the expected $m
  \propto Q^3/T^2$ behavior, showing only limited agreement.  (c,d)
  Data replotted from the study of Ref.~\onlinecite{lago}.  (c)
  300\,K, filled points for a TF of 7.5\,mT, open
  symbols for a longitudinal field of 5\,mT (which was almost
  identical to zero-field data, but a small longitudinal field was
  used to quench any relaxation due to nuclear spins).  (d)
  Measurements in a longitudinal field of 5\,mT.  Note the reduction
  in initial asymmetry in the 5\,K data due to the largely static spin
  ice state.
}
\end{figure}

One of the most ingenious experimental demonstrations of monopoles
came from the muon-spin rotation ($\mu$SR) study of Bramwell {\sl et
  al.} \cite{bramwell} in which 
Dy$_2$Ti$_2$O$_7$
was subjected to a weak transverse field (TF). At very
low temperature (in a regime such that $\ell_{\rm D}\gg \ell_{\rm
  B}\gtrsim\ell_{\rm T}$) it was argued that the relaxation rate of
the muon precession signal $\lambda(B)$ in a magnetic field $B$ could
be interpreted in terms of a fluctuation rate due to monopoles.  By
analogy with Onsager's theory \cite{onsager}, it was argued that one
would expect $\lambda(B)/\lambda(0)\approx 1+\frac{b}{2}+\cdots$ where
$b=\ell_{\rm T}/\ell_{\rm B}=\mu_0 Q^3B/8\pi k_{\rm B}^2T^2$.  This
allows one to measure the field dependence of the relaxation rate and
then infer the monopole charge $Q$.  Bramwell {\sl et al.} measured an
exponentially damped precession signal, interpreting the form of the
damping in terms of magnetic fluctuations resulting from monopoles.
The analysis involved no free parameters and gave a value in very
close agreement with the theoretical value $Q=4.50\mu_{\rm
  B}$\AA$^{-1}$ \cite{castelnovo2008} [see Fig.~\ref{muondata}(a)].
One prediction of this method is that the experimentally measured
quantity $m={1 \over \lambda(0)} {{\rm d}\lambda(B) \over {\rm
    d}B}=b/2B\propto T^{-2}$.  As shown in Fig.~\ref{muondata}(b),
which replots the data of Ref.~\onlinecite{bramwell}, $m \propto
T^{-2}$ only holds for a small subset of the data over a very
limited range of $T$.  Also, $Q\propto m^{1/3}T^{2/3}$ and it is seen
in Fig.~\ref{muondata}(a) that most of the temperature dependence in
the low-temperature behavior of $Q$ is accounted for by the $T^{2/3}$
factor (present {\it by definition}), so that the closeness of the
agreement with theory may not be so compelling as it first appears.
Moreover because the measured $m(T)$ data [Fig.~\ref{muondata}(b)] are
found to fall on cooling down to $\sim 0.3$\,K, below which they are
approximately constant, a limited intersection with the hypothesized
$m\propto T^{-2}$ curve is not unexpected even if the model is inapplicable.

Furthermore, despite the elegance of the theoretical approach, a
surprising feature of these results was that any signal in a weak TF
could be observed at all.  An earlier $\mu$SR study using
longitudinal-field decoupling \cite{lago} had shown that the field at
the muon site was around 0.5\,T, so that a 2\,mT field (the maximum
field used in Ref.~\onlinecite{bramwell}) would not be expected to
lead to a precession signal.  Some data from Ref.~\onlinecite{lago}
are reproduced in Fig.~\ref{muondata}(c,d).  The relaxation of the
precession signal in a TF at 300\,K is the same as that observed in
zero field [Fig.~\ref{muondata}(c)].  That relaxation rate increases
dramatically as the sample is cooled (following an activated behavior
governed by transitions to/from the first excited state doublet of the
crystal field \cite{lago}). At low $T$ [the 5\,K data are shown in
Fig.~\ref{muondata}(d)] the relaxation is so fast that one can only
measure the slower relaxation of the remaining ``$\frac{1}{3}$-tail''
(due to the component of the muon polarization parallel to the local
field).  These results are consistent with a largely static spin ice
state.

\begin{figure}
\includegraphics[width=8.1cm]{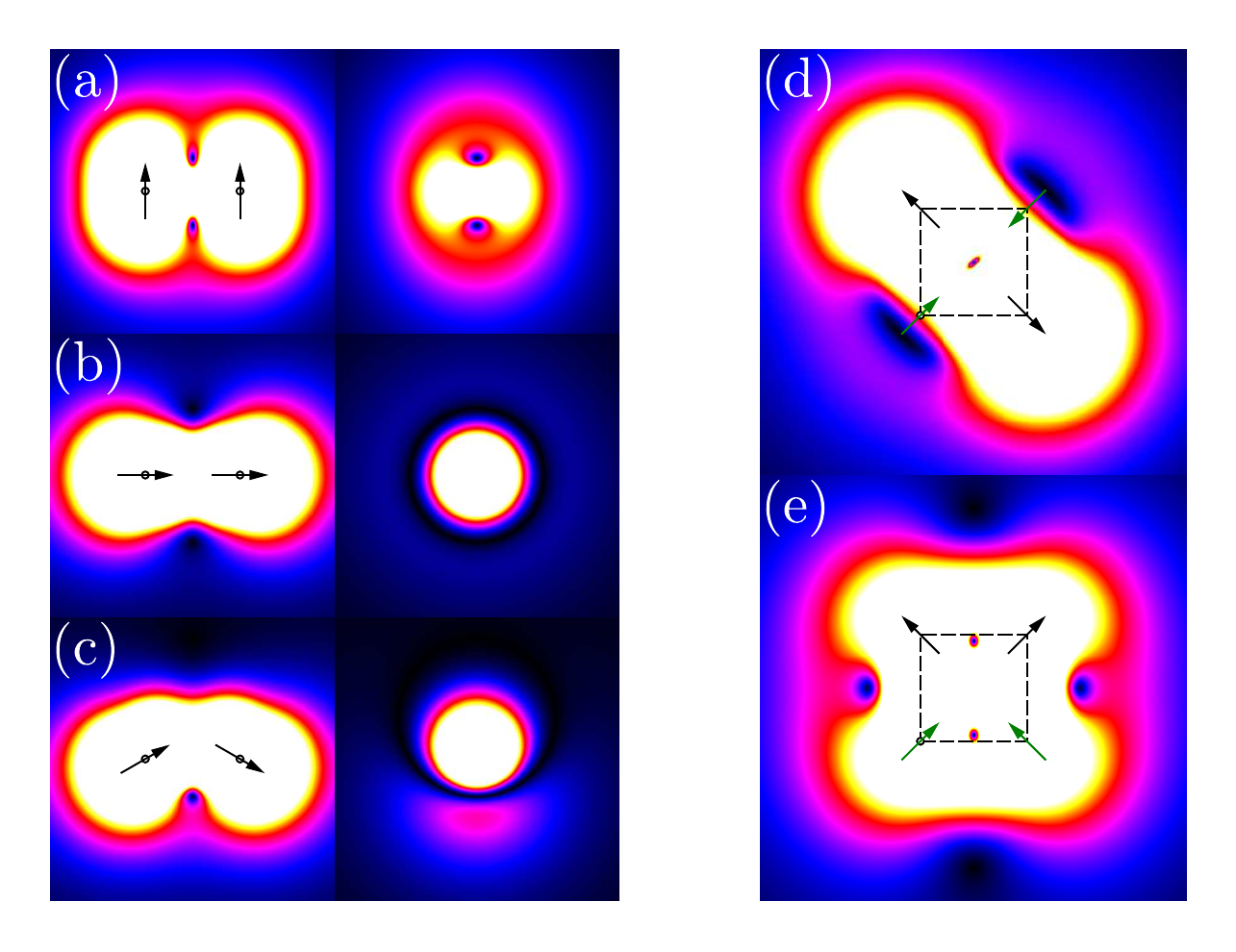}
\caption{\label{dipole} (Color online.)  Distribution of $B_\mu$ in
  the vicinity of (a-c) two spins at positions $(\pm a,0,0)$, showing
  the field in the $y=0$ (left) and $x=0$ (right) planes, and (d-e)
  four spins on the corners of a tetrahedron (inscribed in a cube of
  side length $d$, centred at the origin), satisfying the ice
  rules. (d) $B_\mu$ is shown in the plane $z=0.44d$ with the spins at
  $z=d/2$ ($z=-d/2$) colored black (green).  (e) $B_\mu$ in the plane
  $x=0$.  The areas with the darkest colors have the smallest $B_\mu$.
}
\end{figure}

To be sensitive to a weak TF, the field $B_\mu$ at the muon site needs
to be very low. The hope that there might be some special site of
field cancellation is not borne out by calculations, and the reason
for this can be understood from the following considerations.  First,
it is found that the distribution of the magnitude of $B_\mu$ over the
unit cell volume assuming various ordered arrangements of spins falls
to zero for zero $B_\mu$ \cite{sjb}, so that if present such zeros are
extremely isolated and are unlikely to be obtained by chance.  Even
taking just two spins [say at positions $(\pm a,0,0)$] then the field
zeros occupy either isolated points [e.g.\ if the moments $\parallel
z$, then two zeros are found at $(0,0,\pm a/\sqrt{2})$,
Fig.~\ref{dipole}(a)] or in circles [e.g.\ if the moments are along
$(\cos\alpha,0,\pm\sin\alpha)$, there is a circle of zeros in the $yz$
plane, centred at $(0,0,\frac{3}{2}\tan\alpha)$ with radius
$\sqrt{2+\frac{9}{4}\tan^2\alpha}$, Fig.~\ref{dipole}(b,c)].  For
spins on the corners of a tetrahedron satisfying the spin ice rules
only isolated zeros are found [see Fig.~\ref{dipole}(d,e)].  Crucially
their positions do not share the symmetry of the tetrahedron but
depend on the particular spin ice configuration.  Since muon sites
depend on the electrostatic potential and reflect the crystal
symmetry, then even if a particular muon sits at a site of field
cancellation, many others will sit at crystallographically equivalent
sites in which the field is far from zero.  In a real crystal of
Dy$_2$Ti$_2$O$_7$, containing many tetrahedra, the field from
neighbouring tetrahedra (the spins on which can exist in many
different configurations) will produce additional contributions which
will displace or remove the zeros. In fact, the typical $B_\mu \sim
(\mu_0/4\pi)\mu/r_{\rm d}^3\sim 0.1$\,T, the same order as found in
the earlier $\mu$SR experiment \cite{lago}. This conclusion is in
agreement with other recent calculations \cite{dunsiger,castelnew}.
Thus it is surprising that applying a TF of $\sim 10^{-3}$\,T in the
experiment of Ref.~\onlinecite{bramwell} can have had any effect at
all.  Moreover, the recent experiment of Ref.~\onlinecite{dunsiger}
provides evidence that no such precession signal is in fact observable
when Dy$_2$Ti$_2$O$_7$ is mounted on GaAs and the experiment is
repeated (so that muons missing the sample form muonium and do not
contribute to the observed TF signal).  However, the nature of the
signal observed in Ref.~\onlinecite{bramwell} (with the sample mounted
on silver), which seemingly produces a reasonable estimate of $Q$, has
remained unaddressed.

\begin{figure}
\includegraphics[width=8.1cm]{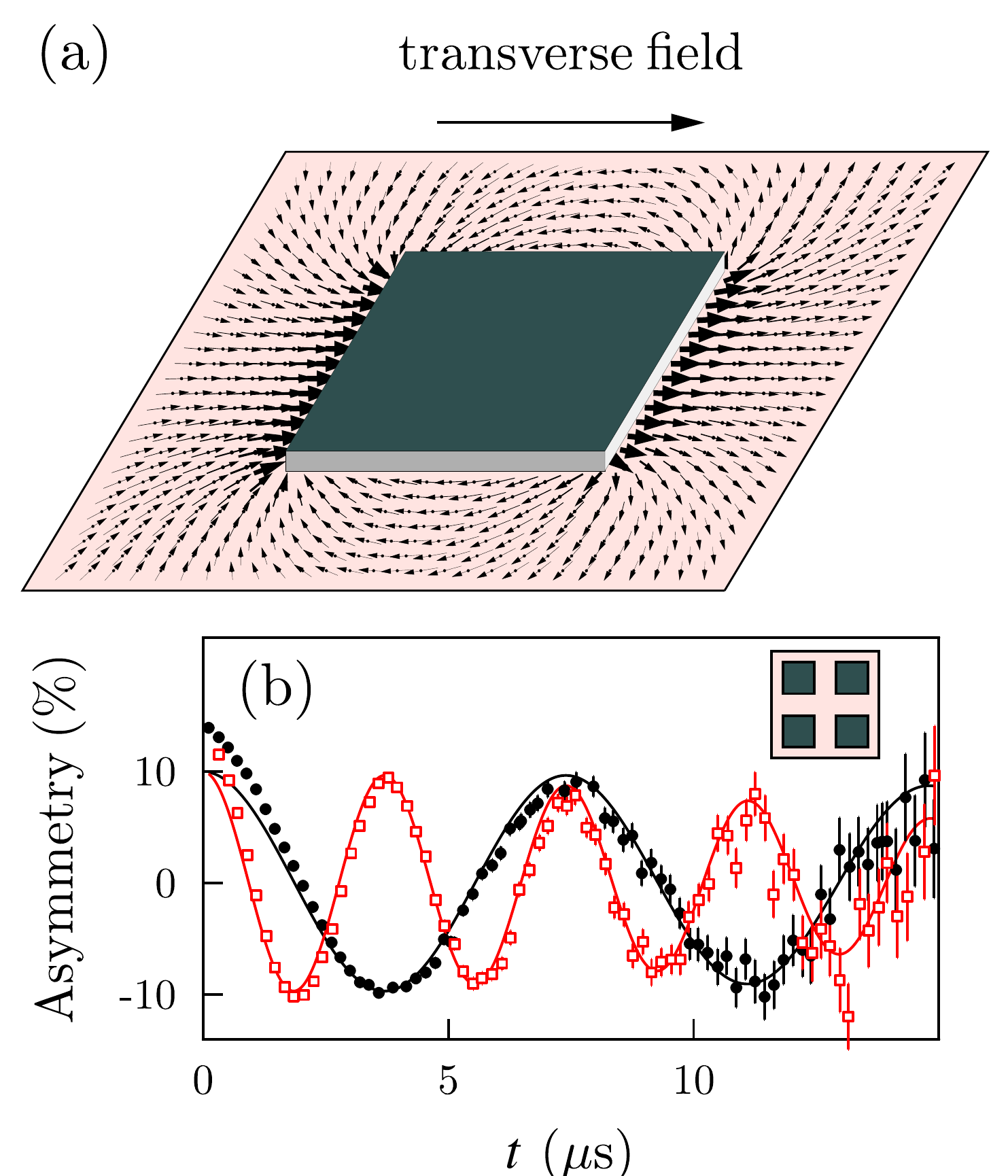}
\caption{\label{3d} (a) The exterior field in the plane of the silver
  plate due to a sample magnetized horizontally and placed on top of
  the plate, evaluated from a closed analytic expression.  The applied
  field adds to this exterior field.  (b) The corresponding muon
  precession signal evaluated for the geometry shown in the inset,
  together with the 100\,mK data from Ref.~\onlinecite{bramwell}.  For this
  geometry, both $\chi$ and the ratio of sample (grey) to sample
  holder (pink) were fitted to the 2\,mT data (red open squares).  The
  fitted geometry is shown to scale and $\chi=0.45$.  The 1\,mT line
  was then evaluated with no further parameter adjustment (data black
  filled circles).  }
\end{figure}

I argue that the most likely resolution of this conundrum is that the
TF muon signal which was observed originates from {\it outside} the
sample (as suspected by Ref.~\onlinecite{dunsiger}), and furthermore
the particular pattern produced by a static, macroscopic exterior
dipolar field produces a muon signal that could be misinterpreted as a
dynamic signal.  Exterior fields are well known to result from
magnetized ferromagnets but are much more unusual in systems with no
long range order.  A macroscopic exterior field directly due to
monopoles is unlikely because $\ell_{\rm D}\ll $sample size (unless
$T<0.2\,$K, but even here $n_{\rm f}$ is likely to be larger than
predicted by DH theory and strongly out of equilibrium
\cite{bramwell2011}, keeping $\ell_{\rm D}$ small) but it is more
likely due to the spin ice magnetization.  When this exterior field
distribution is ``sampled'' by muons implanted in the silver sample
holder over the area close to the sample shown in Fig.~\ref{3d}(a),
simulations show that the resulting relaxation function
[Fig.~\ref{3d}(b)] mimics the exponentially relaxing signal reported
in Ref.~\onlinecite{bramwell}.

\begin{figure}
\includegraphics[width=8.1cm]{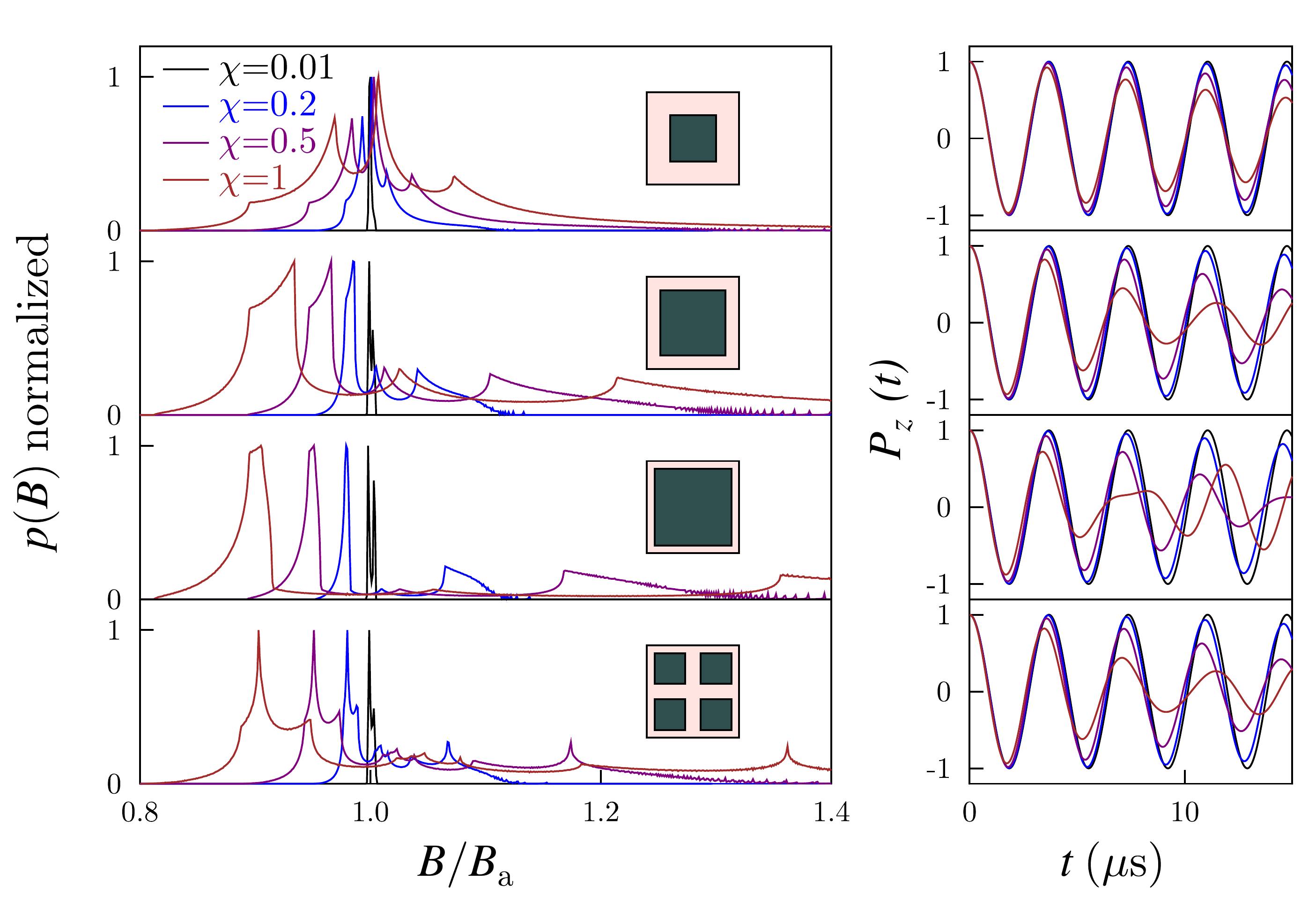}
\caption{\label{multi}
The field distribution $p(B)$ for the muon averaged over the silver plate
for a variety of sample geometries (shown in the insets) and values of $\chi$.  Also shown
are the corresponding simulated muon precession signals $P_z(t)$ for a
field of 2\,mT. 
}
\end{figure}

The simulation of this phenomenon proceeds as follows: In an applied
magnetic field ${\bf B}_{\rm a} = \mu_0 {\bf H}_{\rm a}$, the field
inside and outside the sample is given by ${\bf H}({\bf r}) = (1- \chi
{\cal N}({\bf r})) {\bf H}_{\rm a}$, where $\chi$ is the magnetic
susceptibility and ${\cal N}({\bf r})$ is a symmetrical 3$\times$3
demagnetization tensor that depends on position ${\bf r}$ {\sl even}
outside the sample, which is of particular interest to us.  This
expression is valid under the assumption that the magnetization inside
the sample is uniform.  For a sample with a cuboidal shape an
analytical closed form for ${\cal N}({\bf r})$ can be used
\cite{joseph}.  With the initial polarization along $\hat{\bf n}$, the
resulting time-dependence of the muon polarization can be evaluated
using
\begin{equation}
P_{\hat{\bf n}}(t) = {1 \over A} \int {\rm d}{\bf r} \left(
{B_\parallel({\bf r})^2 \over B({\bf r})^2}
+
{B_\perp({\bf r})^2 \over B({\bf r})^2}
\cos \gamma_\mu B({\bf r})t \right),
\end{equation}
where the integral is over the area $A$ of the silver sample holder
surrounding the sample in which muons are implanted.  Outside the
sample ${\bf B}=\mu_0{\bf H}$ and we have $B_\parallel({\bf r})={\bf
  B} \cdot\hat{\bf n}$ and $B_\perp({\bf r})=\vert{\bf B}({\bf
  r})-B_\parallel({\bf r})\hat{\bf n}\vert$.  Thus if ${\bf B}_{\rm
  a}$ is along $\hat{\bf x}$ and one is measuring $P_z(t)$, then
$B_\parallel({\bf r})=\chi B_{\rm a}{\cal N}_{13}({\bf r})$ and
$B_\perp({\bf r})=B_{\rm a}\sqrt{(1-\chi{\cal N}_{11}({\bf
    r}))^2+\chi^2{\cal N}_{21}({\bf r})^2}$.  The only parameters in
this model are (i) the value of $\chi$ and (ii) the geometry of the
experiment, specifically the size of the region outside the sample
sampled by the muons.  Unless $\chi\gg 1$, the oscillatory component
of $P_z(t)$ dominates and this can be written $\int p(B)\cos\gamma_\mu
B t \,{\rm d}B$ where the function $p(B)$ is plotted in
Fig.~\ref{multi} for various geometries and values of $\chi$, together
with the corresponding $P_z(t)$. The distribution $p(B)$ contains van
Hove singularities of the macroscopic field distribution and broadens
with increasing $\chi$.  It depends on the precise shapes and
arrangements of crystallites on the sample holder, with large density
far from $B_{\rm a}$ due to regions close to the crystallites.  The
fit to data in Fig.~\ref{3d}(b) is quite reasonable (and corresponds
to $\chi=0.45$ and the geometry shown in the inset), though fits with
other geometries are possible.  This value of $\chi$ at 0.1\,K is
larger than observed in zero-field cooled magnetization data
\cite{snyder2004}, but falls short of the equilibrium susceptibility
of spin ice $\chi=\sqrt{3}\pi \ell_{\rm T}/r_{\rm d} \approx 80$
\cite{ryzhkin2011}, showing that the magnetization is still slowly
developing \cite{matsuhira2011} (and this is indeed a manifestation of
the Wien effect \cite{bramwell}).  A single muon measurement at a
particular $B_{\rm a}$ and $T$ is performed of a timescale of $\sim
1$\,hour, so that the effect of slow dynamics (the monopole hop rate
$\sim 10^{3}$\,s$^{-1}$) can become important.  Were larger values of
$\chi$ to be obtained in experiment, the simulations in
Fig.~\ref{multi} show that the TF precession signals should become
strongly distorted by the very strong stray fields, an effect that can
be looked for in future experiments.

The very fast initial relaxation in the data \cite{bramwell}, not
fitted in our formalism [see first 1\,$\mu$s in Fig.~\ref{3d}(b)] is
due to spin dynamics from muons stopped in the sample \cite{lago},
superimposed on the weak TF signal.  The motion of a monopole is
accompanied by the reversal of a Dy spin, resulting in a change
$\Delta B_\mu$ of up to $(2\mu_0/4\pi)2\mu/r^3$, where $r$ is the
distance from the Dy moment to the muon.  Within a radius of $\approx
50$\,\AA\ around the muon there are $\sim 10^4$ Dy ions, each of which
can produce a $\Delta B_\mu$ from tenths to tens of mT.  Thus even
though during the $\mu$SR measurement timescale ($\approx 20$\,$\mu$s)
most monopoles appear static (and the muon experiences a largely
frozen field distribution as discussed above), because the monopole
hop rate is only $\sim 10^3$\,s$^{-1}$\cite{jaubert,snyder2004}, the
muon is coupled sufficiently strongly to such a large number of Dy
ions that even an occasional monopole hop will be enough to contribute
to measurable relaxation in the $\frac{1}{3}$-tail.  The inferred
fluctuation rate of $\sim 10^6$\,s$^{-1}$ \cite{lago} from this
relaxation is thus entirely consistent with this picture.  Moreover
the muon relaxation rate would be expected to track the underlying
monople dynamics (thus the plateau observed in the relaxation time
$\tau$ from a.c.\ susceptibility \cite{snyder2004} is found also in
$\mu$SR \cite{lago}, albeit at a scaled rate).  Below 1\,K however,
$n_{\rm f}$ drops dramatically [see Fig.~\ref{lengths-muons}(a)] and
$\tau$ diverges, while the muon relaxation rate stays approximately
constant \cite{lago,dunsiger,jmmm}, an example of the phenomenon of
persistent spin dynamics observed in many frustrated systems
\cite{mcclarty}.  To understand this effect in Dy$_2$Ti$_2$O$_7$ it is
worth remembering that the muon itself brings substantial kinetic
energy at implantation.  In the final stage of its thermalization, it
loses energy from several tens of keV via charge exchange cycles
whereby an electron successively adds to and is stripped from the muon
\cite{cox99}.  It is conceivable that this process can nucleate
monopoles in the sample (the system is unable to rapidly transport
heat away and is susceptible to thermal runaway
\cite{quenches,runaway}), and when this muon-induced concentration
exceeds the equilibrium concentration, the monopole-induced muon
relaxation rate will settle at a fixed value.  Such an explanation
could have wider applicability in other frustrated magnets.

I am particularly grateful to the following for helpful and enjoyable
discussions: Steve Bramwell, Claudio Castelnovo, Sean Giblin, Michel
Gingras, Peter Holdsworth, Jorge Lago, Tom Lancaster, Roderich
Moessner and Jorge Quintanilla.  This work is supported by the EPSRC,
UK.

\bibliographystyle{apsrev}

\begin{thebibliography}{14}
\bibitem{harris}
M. J. Harris, S. T. Bramwell, D. F. McMorrow, T. Zeiske and K. W. Godfrey,
Phys. Rev. Lett.{\bf 79}, 2554 (1997).
\bibitem{gingras}
S. T. Bramwell and M. J. P. Gingras, Science {\bf 294}, 1495 (2001).
\bibitem{jaubert2011}
L. D. C. Jaubert and P. C. W. Holdsworth,
J. Phys.: Condens. Matter {\bf 23},  164222,  (2011)
\bibitem{hertog}
B. C. den Hertog and M. J. P. Gingras, Phys. Rev. Lett.
{\bf 84}, 3430 (2000);
B. C. den Hertog and M. J. P. Gingras, Can. J. Phys.
{\bf 79}, 1339 (2001).
\bibitem{isakov}
S. V. Isakov, R. Moessner and S. L. Sondhi,
Phys. Rev. Lett. {\bf 95}, 217201 (2005).
\bibitem{pauling}
L. Pauling,  J. Am. Chem. Soc. {\bf 57}, 2680 (1935).
\bibitem{ramirez}
A. P. Ramirez, A. Hayashi, R. J. Cava, R. Siddharthan, and
B. S. Shastry, Nature (London) {\bf 399}, 333 (1999).
\bibitem{ryzhkin}
I. A. Ryzhkin,
JETP {\bf 101}, 481 (2005).

\bibitem{castelnovo2008}
C. Castelnovo, R. Moessner, and S. L. Sondhi, Nature (London) 451, 42
(2008).

\bibitem{castelnovo}
C. Castelnovo, R. Moessner, and S. L. Sondhi,
Phys. Rev. B {\bf 84}, 144435 (2011).


\bibitem{bramwell}
S. T. Bramwell, S. R. Giblin, S. Calder, R. Aldus, D. Prabhakaran,
and T. Fennell, Nature (London) {\bf 461}, 956 (2009).


\bibitem{giblin}
S. R. Giblin, S. T. Bramwell, P. C. W. Holdsworth, D. Prabhakaran,
and I. Terry, Nature Physics {\bf 7}, 252 (2011).

\bibitem{onsager}
L. Onsager, J. Chem. Phys. {\bf 2}, 599 (1934).

\bibitem{lago}
J. Lago, S. J. Blundell, and C. Baines, J. Phys.: Condens. Matter {\bf 19},  326210,  (2007)

\bibitem{sjb}
S. J. Blundell, Phil. Trans. R. Soc. Lond. {\bf A357}, 2923 (1999);
S. J. Blundell, Physica B {\bf 404}, 581 (2009).

\bibitem{dunsiger}
S. R. Dunsiger {\sl et al.}
Phys. Rev. Lett. {\bf 107}, 207207 (2011)
\bibitem{castelnew}
G. Sala, C. Castelnovo, R. Moesnner, S. L. Sondhi, T. Kitagawa,
R. Higashinaka, and Y. Maeno, unpublished.

\bibitem{bramwell2011}
S. T. Bramwell,
J. Phys.: Condens. Matter {\bf 23},  112201,  (2011)

\bibitem{joseph}
R. I. Joseph and E. Schloemann, J. Appl. Phys. {\bf 36}, 1579 (1965).

\bibitem{snyder2004}
J. Snyder, B. G. Ueland, J. S. Slusky, H. Karunadasa, R. J. Cava, and
P. Schiffer, Phys. Rev. B {\bf 69}, 064414 (2004).

\bibitem{ryzhkin2011}
I. A. Ryzhkin and M. I. Ryzhkin,
JETP {\bf 93}, 384 (2011).

\bibitem{matsuhira2011}
K. Matsuhira, C. Paulsen, E. Lhotel, C. Sekine, Z. Hiroi and
S. Takagi,
J. Phys. Soc. Jpn. {\bf 80}, 123711 (2011).

\bibitem{jaubert}
L. D. C. Jaubert and P. C. W. Holdsworth,
Nature Physics {\bf 5}, 258 (2009).

\bibitem{jmmm}
M. J. Harris, S. T. Bramwell, T. Zeiske, D. F. McMorrow, and P. J. C. King
J. Mag. Mag. Mat. {\bf 177}, 757 (1998)

\bibitem{mcclarty}
P. A. McClarty, J. N. Cosman, A. G. Del Maestro, and 
M. J. P. Gingras, J. Phys. Condens. Matter {\bf 23}, 164216
(2011).

\bibitem{cox99}
S. F. J. Cox,
Rep. Prog. Phys. {\bf 72}, 116501 (2009).

\bibitem{quenches}
C. Castelnovo, R. Moessner, and S. L. Sondhi,
Phys. Rev. Lett. {\bf 104}, 107201 (2010).

\bibitem{runaway}
D. Slobinsky, C. Castelnovo, R. A. Borzi,
A. S. Gibbs, A. P. Mackenzie, R. Moessner, and S. A. Grigera,
Phys. Rev. Lett. {\bf 105}, 267205 (2010).
\end{thebibliography}

\end{document}